\documentclass[pre,twocolumn,twoside,byrevtex,superscriptaddress,floatfix]{revtex4-1}

\usepackage{settings}

\setboolean{twocolswitch}{true}
\begin{document}

\title{\protect The Resume Paradox: Greater Language Differences, Smaller Pay Gaps}
\author{
\firstname{Joshua R.}
\surname{Minot}
}
\email{joshua.minot@uvm.edu}
\affiliation{
  Computational Story Lab,
  The University of Vermont,
  Burlington, VT 05401.
} 
\affiliation{
  Vermont Complex Systems Center,
  The University of Vermont,
  Burlington, VT 05401.
} 
\affiliation{
  Department of Computer Science,
  The University of Vermont,
  Burlington, VT 05401.
}

\affiliation{
  MassMutual Data Science,
  Amherst, MA 01002.
}

\author{
\firstname{Marc}
\surname{Maier}
}
\affiliation{
  MassMutual Data Science,
  Amherst, MA 01002.
} 

\author{
\firstname{Bradford}
\surname{Demarest}
}

\affiliation{
  Vermont Complex Systems Center,
  The University of Vermont,
  Burlington, VT 05401.
} 
\affiliation{
  Department of Computer Science,
  The University of Vermont,
  Burlington, VT 05401.
}
\affiliation{
  Neurobotics Lab,
  The University of Vermont,
  Burlington, VT 05401.
}

\author{
\firstname{Nicholas}
\surname{Cheney}
}

\affiliation{
  Vermont Complex Systems Center,
  The University of Vermont,
  Burlington, VT 05401.
} 
\affiliation{
  Department of Computer Science,
  The University of Vermont,
  Burlington, VT 05401.
}
\affiliation{
  Neurobotics Lab,
  The University of Vermont,
  Burlington, VT 05401.
}

\author{
  \firstname{Christopher M.}
  \surname{Danforth}
}

\affiliation{
  Computational Story Lab,
  The University of Vermont,
  Burlington, VT 05401.
} 
\affiliation{
  Vermont Complex Systems Center,
  The University of Vermont,
  Burlington, VT 05401.
} 
\affiliation{
  Department of Computer Science,
  The University of Vermont,
  Burlington, VT 05401.
}

\affiliation{
  Department of Mathematics and Statistics,
  The University of Vermont,
  Burlington, VT 05401.
}

\author{
  \firstname{Peter Sheridan}
  \surname{Dodds}
}
\affiliation{
  Computational Story Lab,
  The University of Vermont,
  Burlington, VT 05401.
} 
\affiliation{
  Vermont Complex Systems Center,
  The University of Vermont,
  Burlington, VT 05401.
} 
\affiliation{
  Department of Computer Science,
  The University of Vermont,
  Burlington, VT 05401.
}
\affiliation{
  Department of Mathematics and Statistics,
  The University of Vermont,
  Burlington, VT 05401.
}

\author{
\firstname{Morgan R.}
\surname{Frank}
}
\email{mrfrank@pitt.edu}

\affiliation{
  Vermont Complex Systems Center,
  The University of Vermont,
  Burlington, VT 05401.
} 
\affiliation{
 Department of Informatics and Networked Systems, 
 University of Pittsburgh, 
 Pittsburgh, PA 15260
} 
\affiliation{
 Digital Economy Lab, 
 Stanford University, 
 Stanford, CA 94305
} 
\affiliation{
 Media Laboratory, 
 Massachusetts Institute of Technology, 
 Cambridge, MA 02139
}

\date{\today}

\begin{abstract}
  \protect
  Over the past decade, the gender pay gap has remained steady with women earning 84 cents for every dollar earned by men on average.
Many studies explain this gap through demand-side bias in the labor market represented through employers' job postings.
However, few studies analyze potential bias from the worker supply-side.
Here, we analyze the language in millions of US workers' resumes to investigate how differences in workers' self-representation by gender compare to differences in earnings.
Across US occupations, language differences between male and female resumes correspond to $11\%$ of the variation in gender pay gap.
This suggests that females' resumes that are semantically similar to males' resumes may have greater wage parity.
However, surprisingly, occupations with greater language differences between male and female resumes have lower gender pay gaps.
A doubling of the language difference between female and male resumes results in an annual wage increase of \$2,797 for the average female worker.
This result holds with controls for gender-biases of resume text and we find that per-word bias poorly describes the variance in wage gap. 
The results demonstrate that textual data and self-representation are valuable factors for improving worker representations and understanding employment inequities.  
\end{abstract}


\maketitle

\section{Introduction}\label{sec:introduction} 
In 2020, the median hourly earnings for women were 84\% of men's earnings in the US \cite{barroso2021gender}. 
Despite a decreasing pay gap in the 1980s and 90s, progress has slowed in recent years. 
Between the 1980s and the 2010s, the portion of the wage gap explained by experience, industry, education, and unionization has decreased---leaving labor economists seeking alternative variables to explain the remaining wage gap \cite{blau2017gender,sullivan2007changing}.
Examples include systemic devaluation of female-dominated fields \cite{levanon2009occupational}, women taking on more unpaid care work than men \cite{ferrant2014unpaid}, and women making more professional sacrifices for family reasons \cite{dunn2018chooses, barroso2021gender, morgan2021unequal}. 

The gender pay gap is typically discussed in aggregate without focusing on specific industries or occupations. 
However, this approach can miss nuanced trends (e.g., within a given occupation).
For example, gender representation in the worker pool varies depending on the occupation, social norms~\cite{napp2022stereotype}, and gender representation among college graduates~\cite{schwartz2022towards}.
But even college majors can be poor abstractions with which to study gender representation.
Many science, technology, engineering, and mathematics (STEM) fields, including computer science (CS), have a larger proportion of male graduates than female on aggregate~\cite{blau2017gender} despite the presence of subfields, such as the human-computer interaction within CS~\cite{cheryan2017some}, where females outnumber males.
Accounting for these differences across contexts requires us to analyze comparable male and female workers engaged in the same type of work with similar educational backgrounds.

How do these factors impact the hiring process?
Several studies investigate employers' hiring decisions to capture gender bias from the demand-side of the labor market. 
For example, employers use automated application systems that can perpetuate systemic bias~\cite{Monedero2019what, ajunwa2016hiring, voxArtificialIntelligence, deshpande2020mitigating}.
A study of job sites Indeed.com and Monster.com found that men receive higher resume rankings \cite{chen2018investigating}.
However, relatively little is known about the bias introduced from the worker supply-side.
Workers provide written texts, such as resumes and cover letters, that are central to the hiring process and provide the opportunity for workers to self-represent their skills, experience, and previous responsibilities.
Accordingly, the same systemic biases on the demand-side of the job market may influence hiring and wages from the supply-side too.
A well-publicized example arose in 2015 when Amazon realized that their resume screening tool was disadvantaging female applicants. 
By developing a system to identify applicants that looked like their (largely male) vision of successful employees, their screening tool ended up disadvantaging women---penalizing resumes with women's colleges and even the word ``women'' \cite{dastinamazon}.
Amazon's screening tool was picking up on stark differences between candidate pools, but more subtle differences undoubtedly exist as well.

Semantic differences in female and male resumes can result from many possible sources.
In general, factors including location and gender correlate with linguistic patterns~\cite{wieling2011quantitative,abitbol2018socioeconomic,schler2006effects, johannsen2015cross,schwartz2013personality}.
As a more specific example, resumes for female MBA students were less likely than their male counterparts to fill in free-text summary and job description fields \cite{altenburger2017there}. 
However, there are relatively few large-scale studies detailing how resume content relates to worker compensation or hiring decisions.
Small case studies suggest that male recruiters can perceive female job seekers as less competent~\cite{dipboye1975relative,cole2004interaction}.

To compare the textual content of resumes, we might turn to existing language models developed to study employer demand-side dynamics through job postings.
Job postings detail professionals' wages and firm productivity \cite{deming2018skill} with language models capable of explaining some of the variance in offered wages \cite{bana2021job2vec}.
For example, gender bias as it relates to job titles is a common theme in AI fairness research \cite{bolukbasi2016man, caliskan2017semantics}---for instance when predicting job and pronoun associations \cite{zhao2018gender}. 
However, the language models developed to study job postings may inherently contain their own biases---stemming from generalized training data or domain specific data sources. 
For instance, the language biases in job ads can have such an impact on recruiting that companies are now offering services to make the text of job ads gender-neutral \cite{miller_job_2017} and reduce gender bias in hiring~\cite{lewis2018will}.
Gender biases in language models mirror societal biases on gender representation in specific jobs \cite{garg2018word, bolukbasi2016man, zhao2018gender, rudinger-etal-2018-gender, sheng-etal-2019-woman, kirk2021bias}.
The gender bias in word embedding spaces can correlate with the gender ratings of professions by human reviewers \cite{grand_semantic_2022}.  
More generally, gender information can improve performance on natural language processing (NLP) tasks utilizing word-embeddings \cite{hovy2015demographic}---highlighting the tension between bias mitigation and performance maximization. 
Thus, a separate approach is required to avoid similar issues in an analysis of resumes.

In the remainder of this study, we quantify the language differences between female and male workers using millions of resumes from US workers.
We relate these differences to broader labor market conditions, such as the gender pay gap and gender representation within specific jobs. 
To do this we combine a large dataset of resumes with demographic and labor economics data. 
We run regression analyses to investigate which language differences between female and male workers describe the gender pay gap in the US labor market. 
In Sec.~\ref{sec:methods} we introduce our datasets and describe our analytical methods. 
In Sec.~\ref{sec:results} we present our results, describing differences in female and male language usage and their association with the gender pay gap. 
Finally, in Sec.~\ref{sec:conclusion} we summarize our findings and suggest directions for future research.

\section{Methods}\label{sec:methods}
\subsection{Language divergence}

\begin{figure*}
    \centering
    \includegraphics[width=\textwidth]{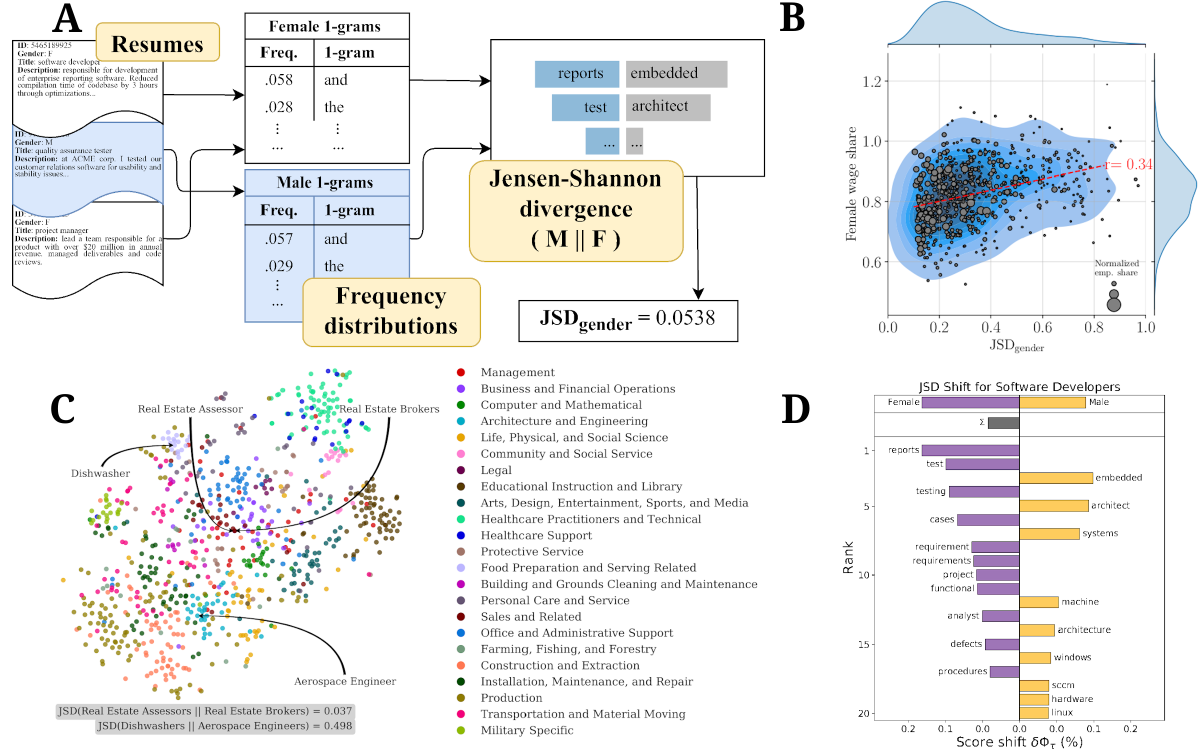}
    \caption{
    \textbf{Female workers earn higher wages when their resumes contain different language than their male counterparts.}
    (\textbf{A}) Given an occupation's resumes, we calculate the 1-gram distributions for female and male workers and the Jensen-Shannon divergence of these two distributions.
    (\textbf{B}) Using wage and resume data from 2005 through 2017, female wage share is larger in occupations with greater language divergence (i.e., females' resumes used different words from males).
    (\textbf{C}) tSNE visualization generated with the pairwise Jensen-Shannon divergence between pairs of detailed occupations in our dataset (i.e., ignoring gender). 
    In this space, ``Real Estate Assessors'' and ``Real Estate Brokers'' use the most similar language, while ``Dishwashers'' and ``Aerospace Engineers'' are most different.
    (\textbf{D}) Explaining language divergence for the resumes of male and female Software Developers.
    Bar size is proportionate to that 1-gram's contribution to overall Jensen-Shannon divergence between female and male distributions (see Fig.~\ref{fig:jsd_shift_sotware_full} for full divergence shift).}
    \label{fig:visual_abstract}
\end{figure*}

We use  Jensen-Shannon divergence (JSD) to quantify the differences in language distributions between female and male workers' resumes. 
JSD compares probability distributions based on their difference from the average of the the distributions.
In our context, the two distributions are the word counts in females' resumes compared to the distribution of words counts in males' resumes within a given occupation.
To better characterize the differences between language distributions, we present word-level contributions to overall divergence values. 
Word-level divergence contribution values allow us to highlight some the most salient differences in language usage by female and male applicants based on empirical observations of language use frequency (see Figure~\ref{fig:visual_abstract}D for an example of JSD shift).
Divergence measurements in general are helpful both for identifying specific words or phrases that have meaningfully different usage between language distributions, as well as summarizing the overall divergence between two distributions.
In our regression analysis we use the total divergence value as an independent variable. 
See Section~\ref{sec:appendix_JSD} for a detailed description of divergence calculations.

\subsection{Data}

We combine a large collection of resumes from the US labor market with population level statistics from the US Bureau of Labor Statistics (BLS) to conduct the current study. 
We also use data from the Occupational Information Network (O*NET) to connect work place activities with specific occupations.  

Throughout the study, we use the BLS 2018 Standard Occupation Classification (SOC) system as a consistent taxonomy of job titles \cite{blsStandardOccupational}. 
The SOC system provides a widely adopted standard for describing and grouping occupations that enables us to aggregate and compare results in a principled fashion. 
Within the SOC there are major, minor, broad, and detailed occupations---with detailed occupations representing the lowest level of the taxonomy.

To match job titles from the resume data to detailed occupations, we use the example titles included with the SOC that correspond to each detailed position to generate word-embeddings for these positions. 
We then use the detailed occupation word-embeddings for matching worker-provided job titles to one of the 867 detailed occupations for further analysis (see Sec.~\ref{subsec:jobtitleclass} for details). 
See Figure~\ref{fig:tsnesbertexamples} for a visual summary of the example title embedding space and list of major occupation categories.

The Occupational Information Network (O*NET) provides detailed work activities (DWAs) that consist of sentence-level descriptions of typical activities performed within given job families \cite{onetNationalCenter}. 
Using a crosswalk between BLS SOCs and O*NET job classes we create sets of DWAs associated with each detailed occupation---these collections of DWAs are used to calculate sentence-embeddings that then generate DWA clusters representative of each occupation. 

Resume data is provided by FutureFit AI and includes machine readable information on over 200 million resumes from the US labor market. 
In the resume dataset we have information on (inferred) worker gender, education, location, and self-reported positions held over the course of their career. 
For each position on a resume, there are fields for dates of tenure, job title, location, company, and a free text description of the position.  

For wage data, we use BLS weekly median earnings collected as part of the current population survey \cite{blsTables}. 
The weekly earnings data provides wage information by detailed occupation and gender for workers in the US labor market.

\section{Results and Discussion}\label{sec:results}
\subsection{Language differences and the gender pay gap}

\begin{figure*}[t!]
    \centering
   \includegraphics[width=\textwidth]{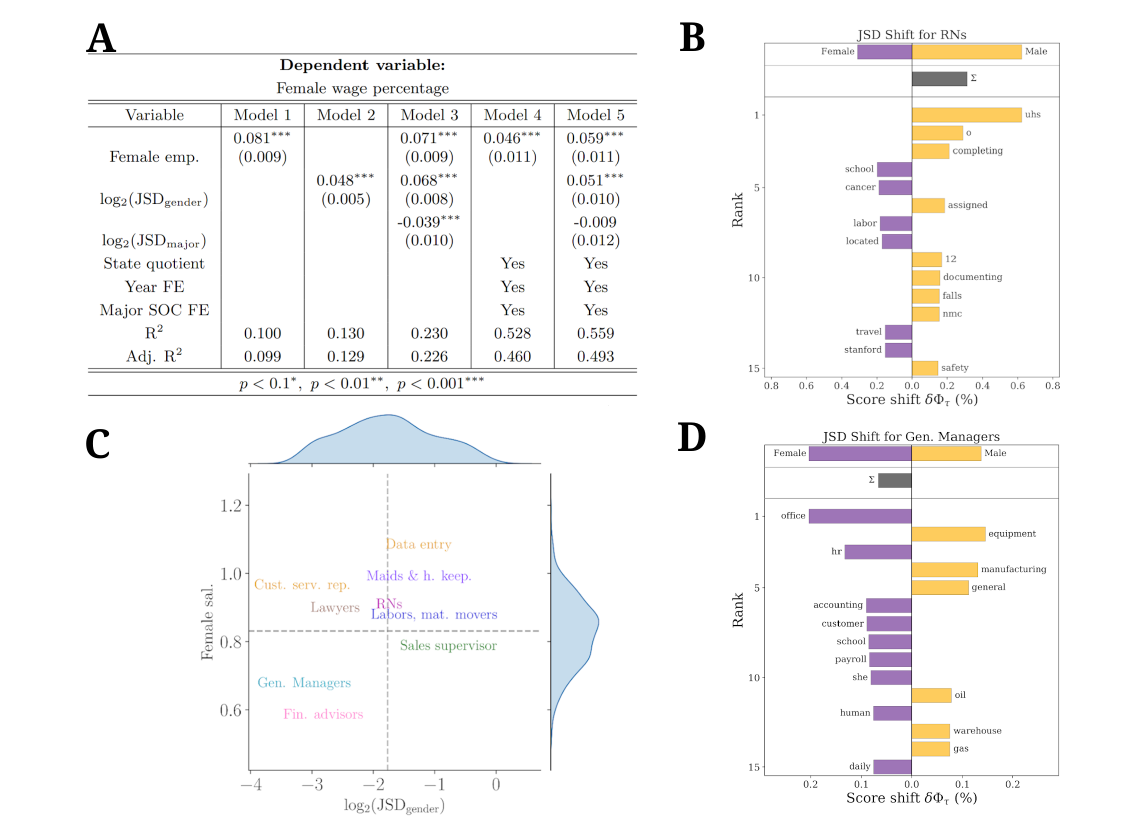}
    \caption{
    \textbf{Female workers earn higher wages when their resumes contain different language than their male counterparts.}
    (\textbf{A}) Summaries for ordinary least squares models including coefficient values and standard errors in parentheses. 
    Language differences capture separate information from female employment share even after controlling for year, state quotient, and occupation type.
    (\textbf{B}) Divergence results for registered nurses in a single year. Registered nurses are a case of a detailed occupation with relatively high language divergence and high female wage share. 
    (\textbf{C}) Quadrant plot showing example job titles relative to the occupations female salary share and JSD between female and male resume 1-gram distributions. Histograms show marginal distributions for full dataset and dashed lines are median values.  
    (\textbf{D}) Divergence results for general managers in a single year. General managers are a case of a detailed occupation with relatively low language divergence and low female wage share. 
    }
    \label{fig:model_composite}
\end{figure*}

\begin{figure}[h!]
    \centering
 \includegraphics[width=\columnwidth]{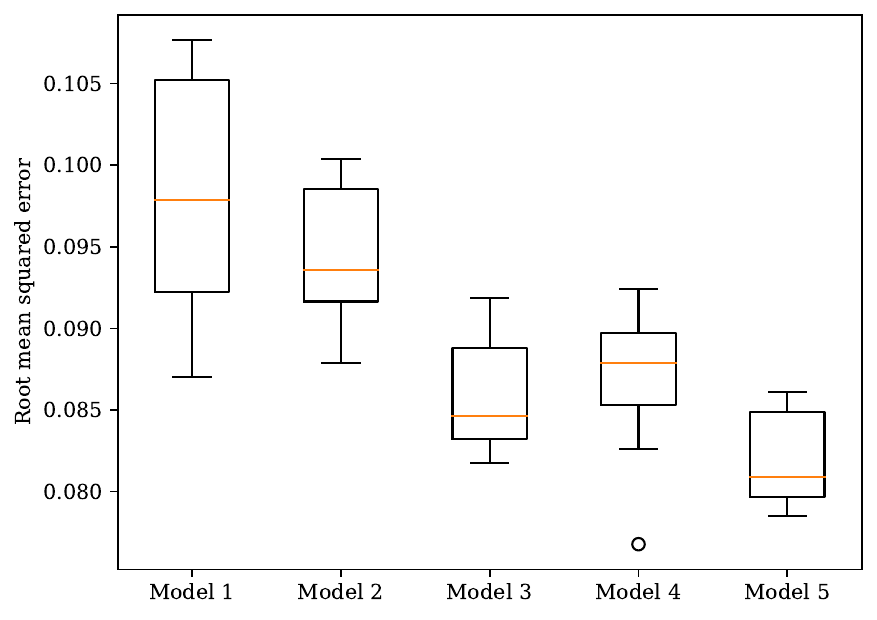}
    \caption{\textbf{Root mean squared error for models predicting women's wage share (female wage percentage)}.
    See table in Fig.~\ref{fig:model_composite}A for model specifications and R$^2$ values.}
    \label{fig:rsme}
\end{figure}

Do males and females with the same occupation use different language to describe their work?
Do language differences correspond to wage differences?
Among workers of the same occupation, we quantify the gender pay gap using women's wages as a percentage of male wages according to $G_f = W_f / W_m$ where $W_f$ and $W_m$ are the median weekly wages for women and men, respectively.  
We include observations of detailed occupations for the 12 years spanning from 2005 to 2017 (see Section~\ref{sec:model_features} for full description). 
According to a simple correlation, females hold a larger wage share in occupations with greater language differences between theresumes of female and male workers ($r=0.34, p<0.001$, see Figure~\ref{fig:visual_abstract}B).
However, this approach ignores several potential confounds including employment share by gender, social differences among occupation types, and changes from year to year.
We control for these additional factors in Figure~\ref{fig:model_composite}A and again find greater wage parity in occupations where female resumes exhibit a greater language difference from males.
For example, language divergence explains a larger share of $G_f$ ($R^2=0.130$) than the female employment share within an occupation ($R^2=0.100$, see Models 1 and 2).
This effect holds even after controlling for state quotient, year, and occupation type where including language divergence increases predictions from an adjusted $R^2$ of 0.460 to 0.493 (see Models 4 and 5).
Further, across all analyses, greater language divergence is significantly positively associated with greater wage parity (i.e., coefficient estimates are significant at the $p<0.001$ level).
See Figure~\ref{fig:rsme} for a comparison of model root mean squared errors and Table~\ref{tab:sal_share_model} for a similar analysis using alternative measures for language divergence.

How much money is associated with language differences?
Controlling for confounds, doubling an occupation's gender language divergence ($\textnormal{JSD}_{\textnormal{gender}}$) is associated with a 5.1\% increase in $G_f$ (see Model 5). 
To put these results in context, we can compare inter-occupation divergence to intra-occupation gender divergence. 
Using the occupation examples from Figure~\ref{fig:visual_abstract}, a doubling in divergence between occupations is similar to moving from the divergence between aerospace engineers and mechanical engineers to the divergence between aerospace engineers and paramedics. 
Another example of doubling divergence is going from the divergence between real estate agents and real estate appraisers to the divergence between real estate agents and financial services sales agents.

What words comprise the language differences between male and female resumes and what is the role of gendered language?
To control for gendered language, we include a Word Embedding Association Test (WEAT) variable (\texttt{w2v bias}) that describes the bias of 1-grams for the distributions corresponding to each observation (see Section~\ref{sec:weat}). 
On its own, \texttt{w2v bias} accounts for roughly 6\% of the variance in $G_f$, but, after adding the variable to the full model we find the coefficient is pushed to effectively 0 and there is no change in variance explained. 
In contrast, when we fit a model to predict the gender employment gap, \texttt{w2v bias} describes roughly 18\% of the variance by itself. 
In the employment gap model, the \texttt{w2v bias} coefficient remains significant when included in the full model, and improves the $R^2$ value by roughly 0.03.
Qualitatively, the employment gap results are consistent with Caliskan \textit{et al.}~\cite{caliskan2017semantics} who find that there is a positive correlation between word-embedding gender bias of job titles and the gender distribution of a detailed occupation ($r = 0.90$, $p < 0.001$). 
In our case, we find that the word-embedding gender bias of the top divergence-driving words and the worker-gender distribution of occupations is positively correlated as well, but with a smaller effect size ($r = -0.38$, $p<0.001$, see Figure~\ref{fig:w2vempshare}). 
However, variation in $G_f$ is better explained by language divergence ($\textnormal{JSD}_{\textnormal{gender}}$) beyond just differences in gendered words.

\subsection{Similarity of job descriptions to canonical activities}
\label{sec:title_sim}

Since males' and females' resumes for the same occupation use different language, which word distribution is most similar to the occupation's description?
We find notable differences in the semantic similarity of position descriptions on resumes and the corresponding set of detailed work place activities (DWAs) provided by the Bureau of Labor Statistic's O*NET database.
We evaluate this difference by embedding job descriptions from resumes and DWAs using an SBERT model.
Our SBERT model \cite{all_mpnet_base-v2, song2020mpnet} is trained on a corpus of books, Wikipedia articles, news stories, and Reddit comments,  which represents a diverse enough corpus to capture variation among occupations in the US economy. 
The distribution of cosine similarity scores provides an indication of how close job descriptions from resumes are to O*NET DWAs in the semantic space---in theory, higher similarity scores represent worker-provided descriptions that are closer to expected descriptions for the job.
See Section~\ref{sec:skills_embed_method} for more details.

\begin{figure*}[t!]
\centering
    \includegraphics[width=.75\textwidth]{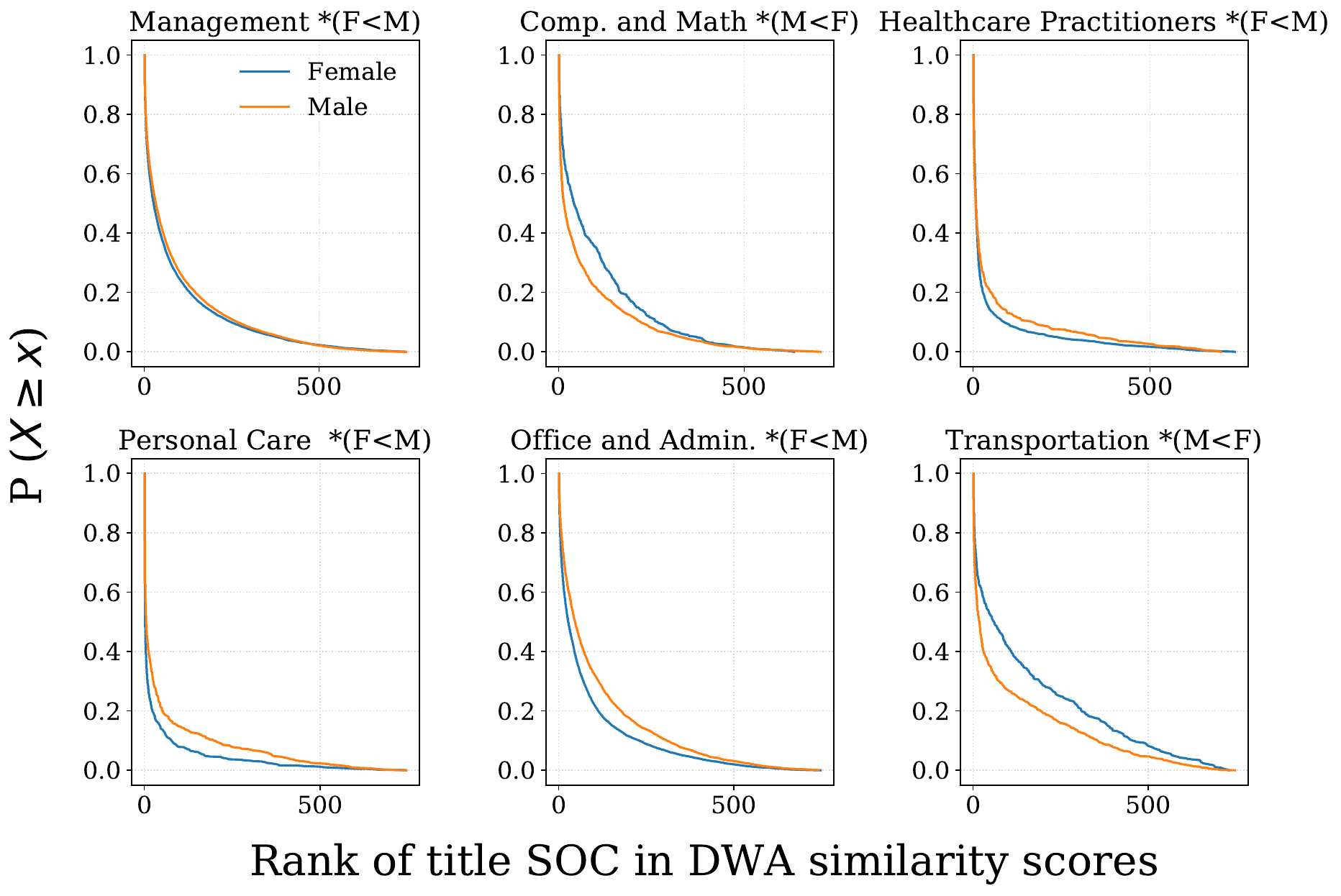}
    \caption{
        \textbf{Women and men's resumes show significant differences in similarity to canonical skills across major occupations}.
        Panels show rank distributions for job titles inferred from resume descriptions matched against detailed workplace activities (DWAs).
        Position descriptions from resumes are embedded using SBERT along with detailed workplace activities pertaining to each detailed SOC (provided by BLS).
        Rank values are for the cosine-similarity scores between DWA and resume embeddings.
        In general, a distribution being shifted to the right indicates that DWAs are less similar to the resumes belonging to that gender. 
        Transportation has a larger difference in the female and male rank distribution than compared with Management. 
        Stars indicate where difference in distribution is significant as detected by a one-sided Kolmogorov-Smirnov test. 
} \label{fig:DWAs_SOC_ranks}
\vfill\hfill

\end{figure*}

Figure~\ref{fig:DWAs_SOC_ranks} shows the rank distribution of true titles matched using the DWA approach for six major occupation classes (i.e., two-digit SOC codes). 
Kolmogorov-Smirnov tests show where there are significant deviations between female and male distributions. 
For instance, in the computer and mathematical occupations male workers on average have their true job titles ranked higher (lower rank value) than females' when using the position descriptions and DWA similarity method. 
Alternatively, for office and administration jobs, females workers' true position is ranked higher than male workers' when using the DWA similarly approach. 
The DWA matching approach provides an indication of where female and male workers may differ in terms of their resume's semantic similarity to canonical descriptions of their jobs. 

\subsection{Resume topics}

\begin{figure*}[t!]
    \centering
    \includegraphics[width=.75\textwidth]{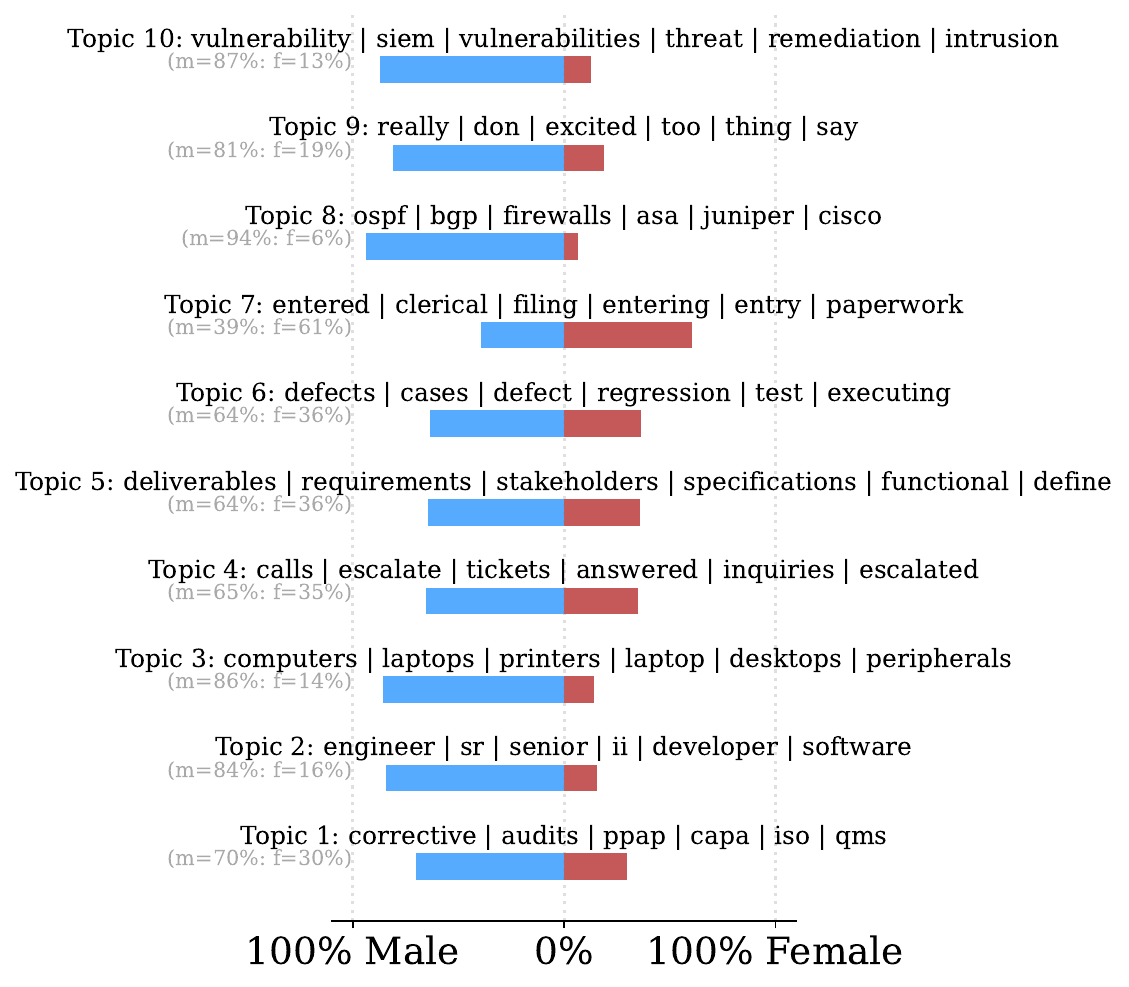}
    \caption{\textbf{Topic models reveal job clusters (and their gender composition) within mathematics and computer occupations}.
    The bars represent the balance of women and men for each topic as determined by assigning resumes to their most similar topic.
    Associated words are the 6 most similar 1-grams in the joint 1-gram and topic embedding space produced by top2vec.
    Topics are ordered from most prevalent (topic~1) to least prevalent. }
    
    \label{fig:top2vecmajor15}
\end{figure*}

So far, we have focused on differences in 1-gram distributions which may undervalue the context in which words are used.
To address this issue, we fit top2vec topic models for a sample of resume position descriptions from each major occupation. 
In Figure~\ref{fig:top2vecmajor15}, we show the top 10 topics for the Computer and Mathematical Occupations major SOC. 
To determine the the gender composition of each topic we first assign positions from resumes to their nearest topic. 
After assigning positions listed on resumes (and associated workers) to each topic, we can calculate the gender balance of workers for the topics. 
For the example topics in Figure~\ref{fig:top2vecmajor15}, we see an overall imbalance towards male workers---with roughly 26\% of Computer and Math positions in our dataset belonging to female workers. 
In topic 10, we can see 1-grams that are associated with cybersecurity such as ``vulnerability'', ``threat'', and ``intrusion.'' 
Topic 10 is also a male-dominated topic with only 13\% of associated resumes belonging to female workers. 
On the other hand, topic 7 contains 1-grams related to clerical work, including words such as ``entered'', ``filing'', and ``paperwork''.
The clerical topic has more female than male resumes, with 61\% of this topic's workers being female. 

\subsection{Gender language divergence}

How do the vocabularies of female and male resumes differ within occupations?
Divergence results for language distributions corresponding to female and male workers within the same detailed occupations reveal some salient differences between genders.
As an example we look at the software developer detailed occupation. 
Figure~\ref{fig:jsd_shift_sotware_full} shows how specific words contribute to the divergence between the language distributions for female and male software developers. 
Overall, we notice an abundance of unique or somewhat rare words in the divergence contributions. 
Terms such ``sccm'', ``gbif'', and ``linkus'' refer to specific tools, organizations, or products.
Indeed, there is a heavy tail of rare terms owing partially to the prevalence of specific tools and technologies in the field. 
We see more common words such as ``requirements'', ``cases'', and ``tests'' appearing more frequently in female resumes, which likely corresponds to a higher prevalence of female developers in the sub-fields of quality assurance \cite{blsCPS2022}. 
On the other hand, we see the words ``embedded'' and ``hardware'' appear more frequently in the male distribution, likely owing to the higher prevalence of men in sub-specialities that deal more directly with computer hardware \cite{blsCPS2022}. 

\begin{figure}
    \centering
\includegraphics[width=.9\columnwidth]{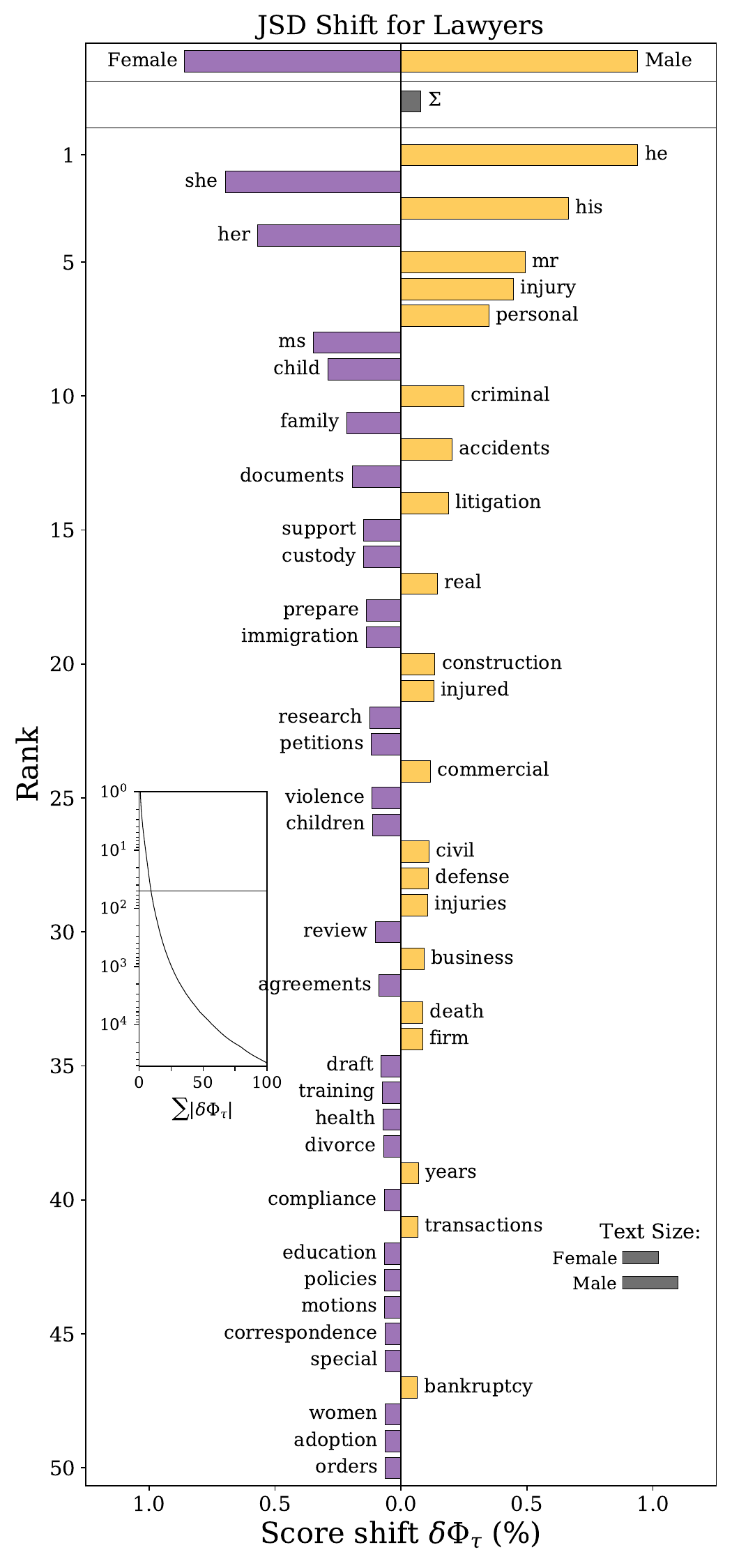}
    \caption{\textbf{Individual word contributions to Jensen-Shannon divergence (JSD) highlight the specific language usage that drives overall divergence values.}
    Figure shows a word shift \cite{gallagher2021generalized} using JSD between female and male lawyers.
    The bar size associated with individual 1-grams is proportionate to that 1-gram's contribution to the overall JSD value. 
    In the case of lawyers, the shift is likely picking up on differences in the legal specialities of each population. 
    The 1-grams driving the female distribution's contribution to overall divergance are indicative of family law (e.g., ``family'', ``child'', and ``divorce''). 
    On the other hand, the 1-grams associated with male lawyers seem associated with personal injury and commercial law (e.g., ``business'', ``transactions'', and ``bankruptcy'').
    }
    \label{fig:jsd_shift_lawyers_full}
\end{figure}

Moving to a different detailed occupation, Figure~\ref{fig:jsd_shift_lawyers_full} provides an example JSD shift from the legal profession where we see less technical words driving the divergence. 
In this case, we see the female distribution contains relatively more occurrences of terms related to family law (e.g., ``family'', ``domestic'', ``guardianship''). 
The distribution for male lawyers contains relatively more words related to intellectual property (e.g., ``intellectual'', ``secrets'') and personal injury (e.g., ``accident'', ``injury''). 

\subsection{Inter-occupation language divergence}

Divergence values in isolation do not lend themselves to intuitive interpretation. 
For instance, what does a JSD value of 0.08 mean relative to real world outcomes?
To contextualize the divergence results presented throughout this study, we calculate the pairwise language distribution divergence for all detailed occupation codes (including workers from both genders). 
In Figure~\ref{fig:visual_abstract}C, we present a tSNE visualization of an embedding space derived from pairwise JSD values. 
The tSNE visualization shows the divergence values embedded in a 2-dimensional space where the grouping of major occupation categories is apparent. 

We select detailed occupations with greater than the median number of resumes in order to focus on more common occupations that are best represented in our dataset. 
Using JSD, the two most similar detailed occupations are \texttt{Appraisers and Assessors of Real Estate} and \texttt{Real Estate Brokers} (JSD = 0.037).
Notably, these occupations are from different major occupation groups, but due to their domain specific experience and skills the JSD measure is relatively small between them. 
The two detailed occupations with the greatest divergence are \texttt{Dishwashers} and \texttt{Aerospace Engineers} (JSD = 0.498).

\section{Concluding Remarks}\label{sec:conclusion}
In the current study, we describe aggregate level language differences between female and male workers' resumes and observe an association between these differences and the gender pay gap within US occupations. 
We show the gender composition of resume topics for a major category of jobs, shedding light on the signal present in the resume-text dataset and its relation to gender (im)balances. 
We also present follow-on work to prior research that expands on the connection between gender bias present in semantic spaces with on-the-ground realities of gender distributions within occupations. 
Taken together these results demonstrate the value in leveraging text data to improve our understandings of workers' skills, experience, and other factors. 
We show that textual data contains signals that help describe real world harms such as the wage gap. 
Further, we see how one must be thoughtful about constructing text-based features---in our case examining wage gap and employment share required two different lenses to surface meaningful associations. 

Do language differences in resumes correspond to earnings differences?
Males earn more than females in expectation for most occupations and that male resumes are semantically more similar to expertly-defined occupation descriptions from the US Bureau of Labor Statistics (see Fig.~\ref{fig:visual_abstract}B and Fig.~\ref{fig:model_composite}A).
Thus, our original hypothesis was that occupations where resumes for female workers used different language than males would be the occupations with the greatest gender pay gap.
However, surprisingly, occupations with higher divergence between female and male language distributions exhibit reduced gender pay gaps. 
The idea that greater differences in language could lead to improved equity in compensation runs counter to intuition.
We theorize that the effect observed may partially owe to sub-specialization of female workers---with greater experience and more skills potentially being required to overcome the pay gap. 
This possibility represents an interesting avenue for future work because of its potential to shape equity practices in hiring and business leadership.
For instance, one study on hiring for managerial positions found that while men may be selected for leadership potential women are selected for leadership performance \cite{player2019overlooked}.
More generally, reviews of US and UK labor surveys found that women are broadly held to stricter performance standards compared with men for the same job \cite{gorman2007we}.

Our regression analysis demonstrates that job-specific language divergence measurements detect salient differences between female and male workers' resumes with respect to the gender wage gap (see Fig.~\ref{fig:model_composite}A). 
These differences are likely more than a simple function of gender employment share or other factors such as location, year, or major occupation.
Indeed, the divergence measures are poor predictors of female employment percentage as seen in Table~\ref{tab:emp_share_model} and the correlation in Fig.~\ref{fig:rtdempshare} ($r=0.01$). 
Further, the divergence measures appear to capture information that is not entirely described by measurements of gender bias in a word-embedding space. 
Taken together, these results suggest that there are latent characteristics of language distributions for female and male workers that help describe at least part of the gender pay gap.

Divergence results provide the first indication of meaningful differences in language usage between genders. 
We show that the individual words contributing to the divergences are informative.
Indeed, viewing individual 1-gram divergence contributions provides explainability of our divergence results. 
Another indication of gender disparities is seen with our matching of canonical workplace activities for specific jobs with text from resume position descriptions. 
The activity matching shows differences for female and male workers---suggesting that the idealized candidate for specific jobs may have gender bias, at least relative to empirically observed resume data. 
Topic modelling provides an indication of how resume text commonly grouped together along with the gender composition of these groupings.

The job-specific language divergence measurements are likely influenced by many factors. 
Broadly, geographic variation in language usage, temporal effects, and industry effects are all factors that likely influence the language characteristics we report here. 
We control for these specific factors using fixed effects for time and major occupation category along with including a control for detailed occupation employment at the state-level.  
There are also factors related to specific firms and educational institutions that may affect the language features of resumes---something we do not control for in the present study. 

We have presented some evidence that the language distribution divergences may be partially driven by sub-specialization (e.g., quality assurance vs. hardware engineers for software developers). 
Indeed, no two workers are exactly the same. 
Experience, qualifications, and skills, all contribute to workers that are best thought of as having high dimensional characteristics.
Language is one manner by which to capture salient information that would be difficult to collect in tabular form---especially given current data repositories. 

Thinking broadly about our results, we believe there is some logic behind the finding that the representation gap is associated with gender bias in the word-embedding space, while the wage gap would be associated with job-specific language differences. 
For one, the gender bias present in language models is likely owing to societal associations of genders and occupations based on the demographics of workers (this point can lead to feedback loops with subconscious selection for stereotypical workers). 
On the other hand, intra-occupation wage gaps are less frequently discussed and likely less central to societal perceptions of jobs and worker characteristics. 
Instead, wage gaps within occupations are more likely related to factors that are also encoded in resumes---skills, experience, certifications, and so on.

Some efforts exist to combat these biases in AI applications.
The National Institute for Standards and Technology (NIST) strives to improve standards for quantifying and understanding bias in AI systems \cite{schwartz2022towards}. 
The main sources of bias in an AI system can be roughly attributed to structural, human, and statistical factors. 
We can conceptualize mitigation as occurring primarily at the dataset level \cite{feldman2015certifying, hitti2019proposed}, at test and evaluation time, and/or with modification of human factors.
In all mitigation approaches it is helpful to understand domain specific bias present in the main data artifacts of the field. 
Ultimately, bias mitigation has ethical and performance implications. 
Improving representations of individuals in AI systems not only improves outcomes with respect to ethical considerations, but also improves our understanding of systems through a scientific lens. 
Indeed, we hope the current work can serve to reduce the precursors to bias in AI systems while also improving how workers are represented in future-of-work and labor economics research. 

When studying data bias, and commenting on its potential relation to fairness, it is important to define our working definitions of bias and harm \cite{blodgett2020language, altman2018harm}.
In the current work we focus on quantifying gender bias present in resume data.
More specifically, in our case gender bias takes the form of \textit{meaningful} differences in the textual data associated with female and male workers---where \textit{meaningful} is in turn informed by the biases' relation to known harms. 
The primary harm we are concerned with is the gender pay gap. 
Female workers earning less than their male counterparts presents tangible harms related to economic hardships and is indicative of issues related to career advancement. 
Secondary (and related) to the wage gap, is the gender representation gap in a given field. 
Aggregate gender representation in a given field does not clearly present the same level of specific harm as wage.
However research suggests gender representation imbalances can be a precursor to wage disparities \cite{levanon2009occupational}. 
Additionally, there are concerns about feedback loops in fields being viewed as stereotypically favoring female or male workers, subsequent discrimination, and trainees having access to peer mentorship.
Finally, seminal studies on the gender bias present in language models
use the proportion of female to male workers in a field as reference point when demonstrating gender biases \cite{bolukbasi2016man,caliskan2017semantics,grand_semantic_2022}.

We were limited to examining binary cases of gender due to the constraints on the available datasets (both the resumes and BLS data are coded with female and male genders). 
Furthermore, we were not able to evaluate the accuracy of the gender variable in our resume dataset---a potential issue given some gender classifiers have been shown to vary in accuracy along racial and ethnic lines \cite{santamaria2018comparison}.  
Our resume dataset does not claim to be a totally representative sample of workers from the US labor market, and the potential for sampling bias influencing our results cannot be ruled out.
Going forward, more research is needed comparing text-as-data approaches to worker representation with conventional labor economic frameworks such as the BLS SOC.  

We are also unable to comment on the direct causes of the wage gap based on our analytical framework---we are merely able to present notable associations. 
Future work could investigate how modifying language in resumes affects job screening and hiring processes---putting our theories to the test in more real world settings. 
Addressing the demand side of the labor market and how job postings interact with workers and the gender pay gap is another necessary step to understanding how worker characteristics and societal biases interact. 
Our study provides an initial, system level analysis of our research questions. 
Given the amount of variation between fields, tailored case studies examining the gender pay gap based on resumes from specific fields is also worthwhile. 

The text-as-data paradigm, when used correctly, can help us create improved representations of workers for the purposes of labor economics and future-of-work research, worker re-skilling, talent screening, and other areas.  
But first we need to understand how linguistic features interact with other worker attributes and may contribute to real world harms. 
The gender wage gap is but one example of these harms---one for which data was available for the present study---but there are other worker attributes that should be understood in relation to textual data (e.g., race, parental status, age, etc.). 
Using free text descriptions could unlock a vast amount of information that is increasingly important in today's modern workforce---especially as jobs, skills, and careers change more rapidly than ever. 
Accessing the promise of this data must come in tandem with understanding how biases and inequalities are contained within---both for the purposes of better understanding and mitigating these issues. 
For instance, other research has emphasized the limitations of efforts to debias language models \cite{gonen-goldberg-2019-lipstick}, and our work suggests that salient gender signals exist outside the dimensions of gender that would commonly be encoded in a general purpose language model.

\acknowledgments 
The authors are grateful for the computing resources provided by the Vermont Advanced Computing Center 
and financial support from the Massachusetts Mutual Life Insurance Company and Google.
This research is supported in part by the University of Pittsburgh Center for Research Computing.
We are also grateful for conversations with John Meluso regarding the current study. 

\bibliography{references}

\clearpage

\newwrite\tempfile
\immediate\openout\tempfile=startsupp.txt
\immediate\write\tempfile{\thepage}
\immediate\closeout\tempfile

\setcounter{page}{1}
\renewcommand{\thepage}{S\arabic{page}}
\renewcommand{\thefigure}{S\arabic{figure}}
\renewcommand{\thetable}{S\arabic{table}}
\setcounter{figure}{0}
\setcounter{table}{0}

\section{Supplementary Information (SI)}
\renewcommand{\thefigure}{S\arabic{figure}}
\renewcommand{\thetable}{S\arabic{table}}

\subsection{Job title classification}
\label{subsec:jobtitleclass}

Using sentence transformers (or SBERT) we create job title word-embeddings, $\overrightarrow{t}$, for all titles appearing in the FutureFit AI resume dataset. 
We apply a similar procedure to the BLS SOC example titles, starting by creating a word-embedding for each title (BLS provides 4-10 example titles for each detailed occupation).
Next, we take the average embedding of the example job titles for each detailed SOC---this mean vector will represent the detailed SOC category for the resume title classification task.
The collection of these mean vectors for occupations, $\overrightarrow{o}$, forms the set of all occupation mean vectors, $O$.
Classification is performed by searching for the most similar SOC embeddings for each resume position title embedding, $\overrightarrow{t}$ in the resume title set $T$. 
Using cosine similarity, we calculate the pairwise cosine similarity between each title and all 867 detailed SOC embeddings. 
We take the \texttt{argmax} of the cosine similarly vector for each resume title, and assign the corresponding SOC job title as a candidate match, 

\[ \argmax_i \textnormal{cos} (\overrightarrow{t}, \overrightarrow{o}_i) \quad , \forall i \]

Finally, we threshold by calculating the empirical cumulative distribution function for all pairwise matches (i.e., not just the `best' or most similar match) and retaining the candidate matches that are greater than $99.9\%$ of all cosine similarity scores.

\begin{figure*}[ht!] 
\includegraphics[width=\textwidth]{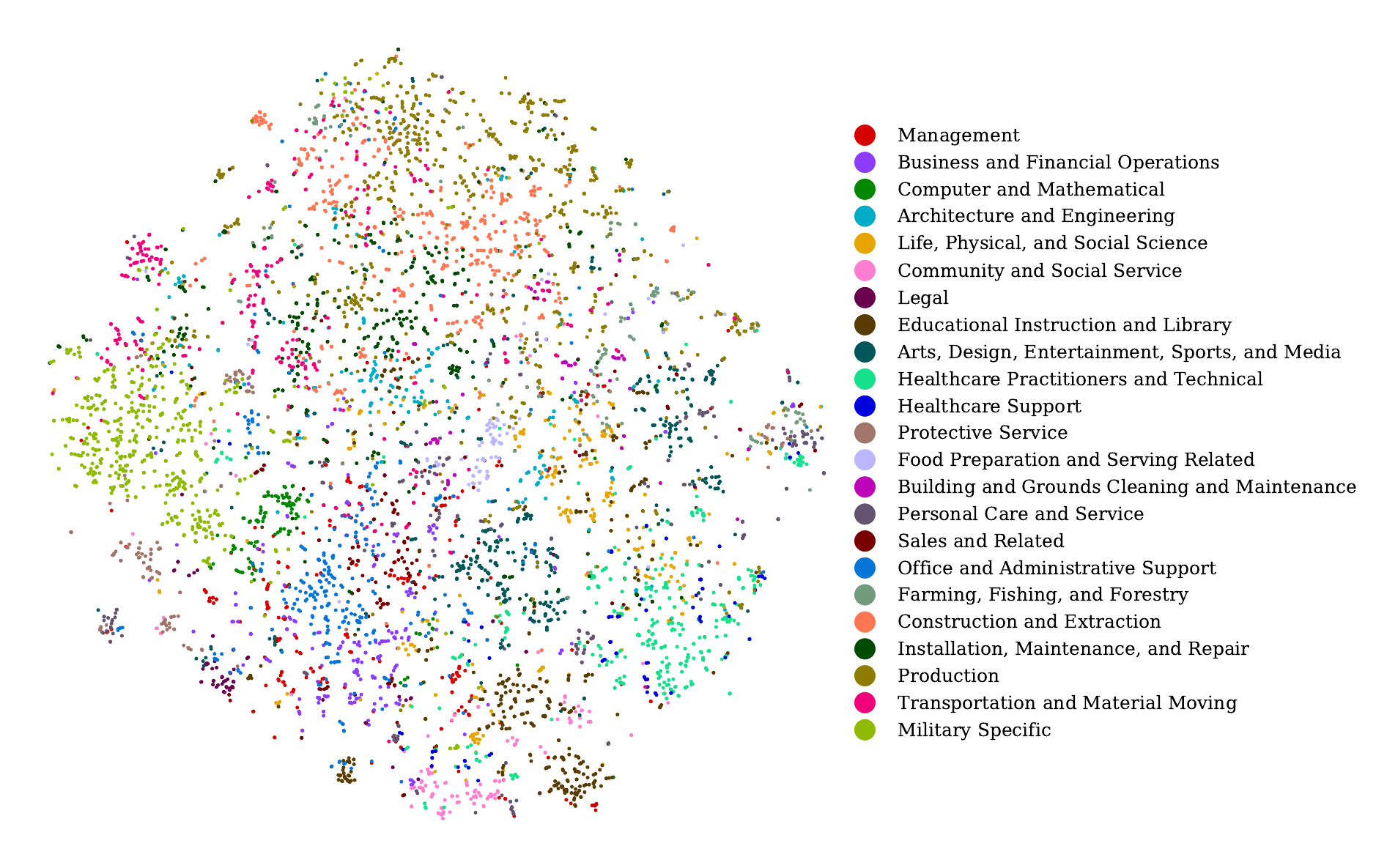}

    \caption{
        \textbf{Embedding space of job titles visually corresponds to salient job characteristics such as major occupation category (shown) and level of educational attainment (Fig.~\ref{fig:tsnesbertexamples_education})}.
        Points produced by t-distributed stochastic neighbor embedding (tSNE) of example job titles using sentence BERT (SBERT).
        The embedding visualization provides a general indication of semantic similarity of job titles in the SBERT semantic space. 
        Points represent an individual example job title provided by the US Bureau of Labor Statistics Standard Occupational Classification (SOC) system. 
        Points are colored based on their membership in one of 23 major occupation categories. 
        See Fig.~\ref{fig:tsnesbertexamples_education} for embedding colored by educational attainment.} \label{fig:tsnesbertexamples}

\end{figure*}

\begin{figure*}[ht!] 
       \includegraphics[width=\textwidth]{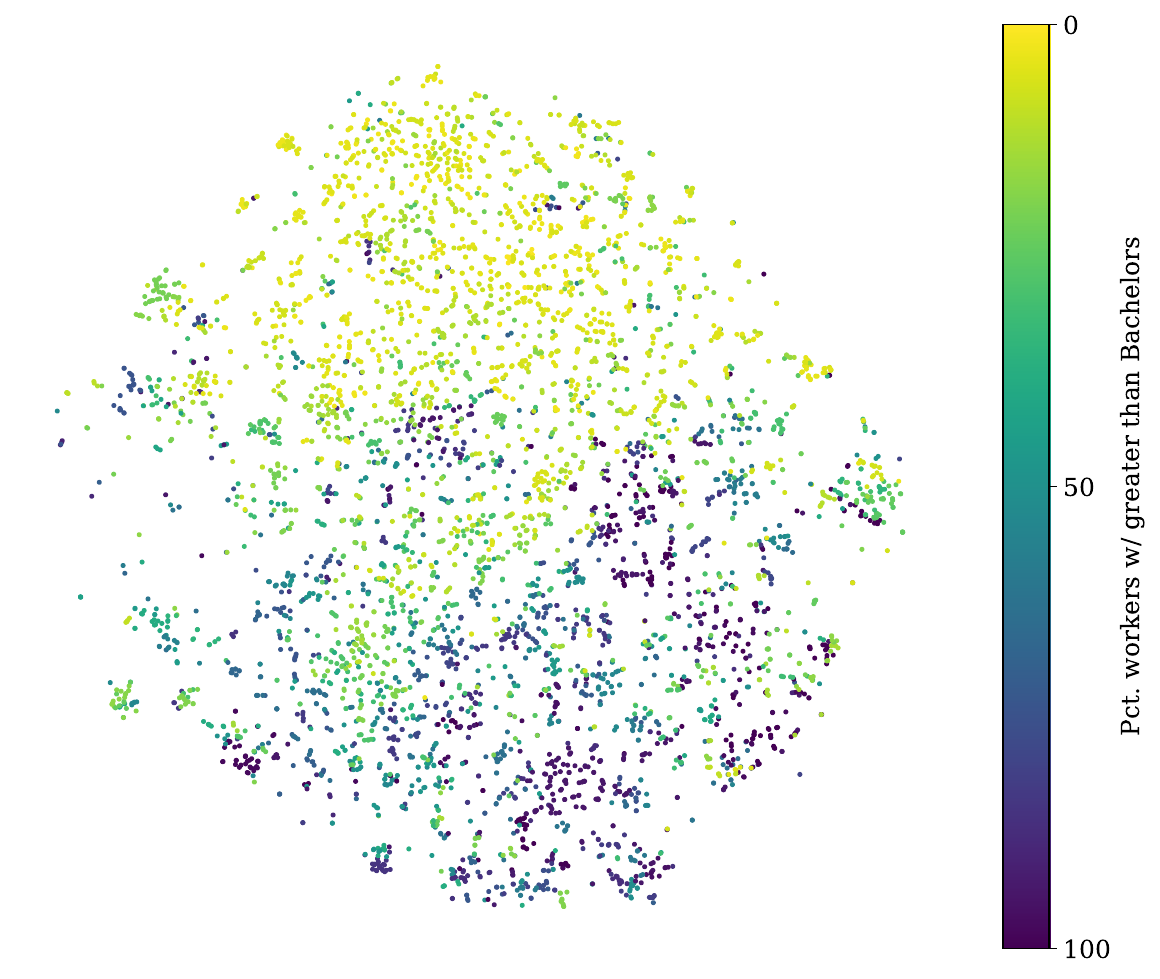}

    \caption{
        \textbf{Embedding space of job titles visually corresponds to level of educational attainment}.
        Points produced by t-distributed stochastic neighbor embedding (tSNE) of example job titles using sentence BERT (SBERT).
        The embedding visualization provides a general indication of semantic similarity of job titles in the SBERT semantic space relative to average educational attainment for each position. 
        Points represent an individual example job title provided by the US Bureau of Labor Statistics Standard Occupational Classification (SOC) system. 
        Points are colored based on the percentage of workers with a bachelors degree or higher (where data is available, note the wholly missing points for Military Specific occupations.).
        }  \label{fig:tsnesbertexamples_education}

\end{figure*}

\clearpage

\subsection{Divergence calculations}
\label{sec:appendix_JSD}

The Jensen-Shannon divergence between two sets, $\Omega_1$ and $\Omega_2$, is calculated as follows, 

\begin{equation}\label{eq:JSD}
\textnormal{JSD}(\Omega_1||\Omega_2) = \frac{1}{2}D(P_1||P_\textnormal{M}) + \frac{1}{2} D(P_2||P_\textnormal{M})
\end{equation}

Where, $D(P_x||P_y)$ is the Kullback-Leibler divergence between probability distributions $P_x$ and $P_y$,
$P_\textnormal{M} = \frac{1}{2} (P_1 + P_2)$,
and $P_i$ is the probability distribution for~set~$\Omega_i$.

To generate the language distributions divergences we start by counting the occurrences of 1-grams (space separated character sequences) in resumes delimited by gender and detailed occupations. 
We then calculate the JSD between female and male language distributions for each occupation.  
For example, for our main analysis we end up with a divergence value for female and male software developers (see Fig.~\ref{fig:visual_abstract}A~and~D). 
For the analysis of differences between genders we mainly compare language within detailed occupations to control for differences in occupation specific language usages and associated gender imbalance. 

We use the \texttt{Shifterator} python package \cite{gallagher2021generalized} to visualize the contributions of individual 1-grams to overall divergence values (see Fig.~\ref{fig:jsd_shift_lawyers_full} for a full JSD shift).
Contributions for individual elements $\tau$ to overall JSD can be calculated by decomposing Eq.~\ref{eq:JSD},
\begin{equation}
    \textnormal{JSD}(\Omega_1||\Omega_2)_\tau = -m_\tau \log m_\tau + \frac{1}{2} \left( p_{\tau,1} \log p_{\tau,1} + p_{\tau,2} \log p_{\tau,2} \right)
\end{equation}
where $m_\tau$ is the probability of element $\tau$ in the mixed distribution $P_\textnormal{M}$ and $p_{\tau,i}$ is the probability element $\tau$ in distribution $P_i$.

\subsubsection{Alternative divergence measurement}
\label{SI:rtd}

In addition to Jensen-Shannon divergence, we use rank-turbulence divergence (RTD) \cite{dodds2020allotaxonometry} to quantify the differences in language distributions for female and male workers.

The rank-turbulence divergence between two sets, $\Omega_1$ and $\Omega_2$, is calculated as follows, 
\begin{equation}
\begin{aligned}
D^{R}_{\alpha}(\Omega_1 || \Omega_2) 
&= \sum \delta D^{R}_{\alpha,\tau}  \\
&= \frac{\alpha +1}{\alpha} \sum_\tau \left| \frac{1}{r_{\tau,1}^\alpha} - \frac{1}{r_{\tau,2}^\alpha} \right| ^{1/(\alpha+1)} \,,
\end{aligned}
\end{equation}
where $r_{\tau,s}$ is the rank of element $\tau$ ($n$-grams in our case) in system $s$ and $\alpha$ is a tunable parameter that affects the impact of starting and ending ranks.

\begin{figure*}[h]
    \centering
    \includegraphics[width=.9\textwidth]{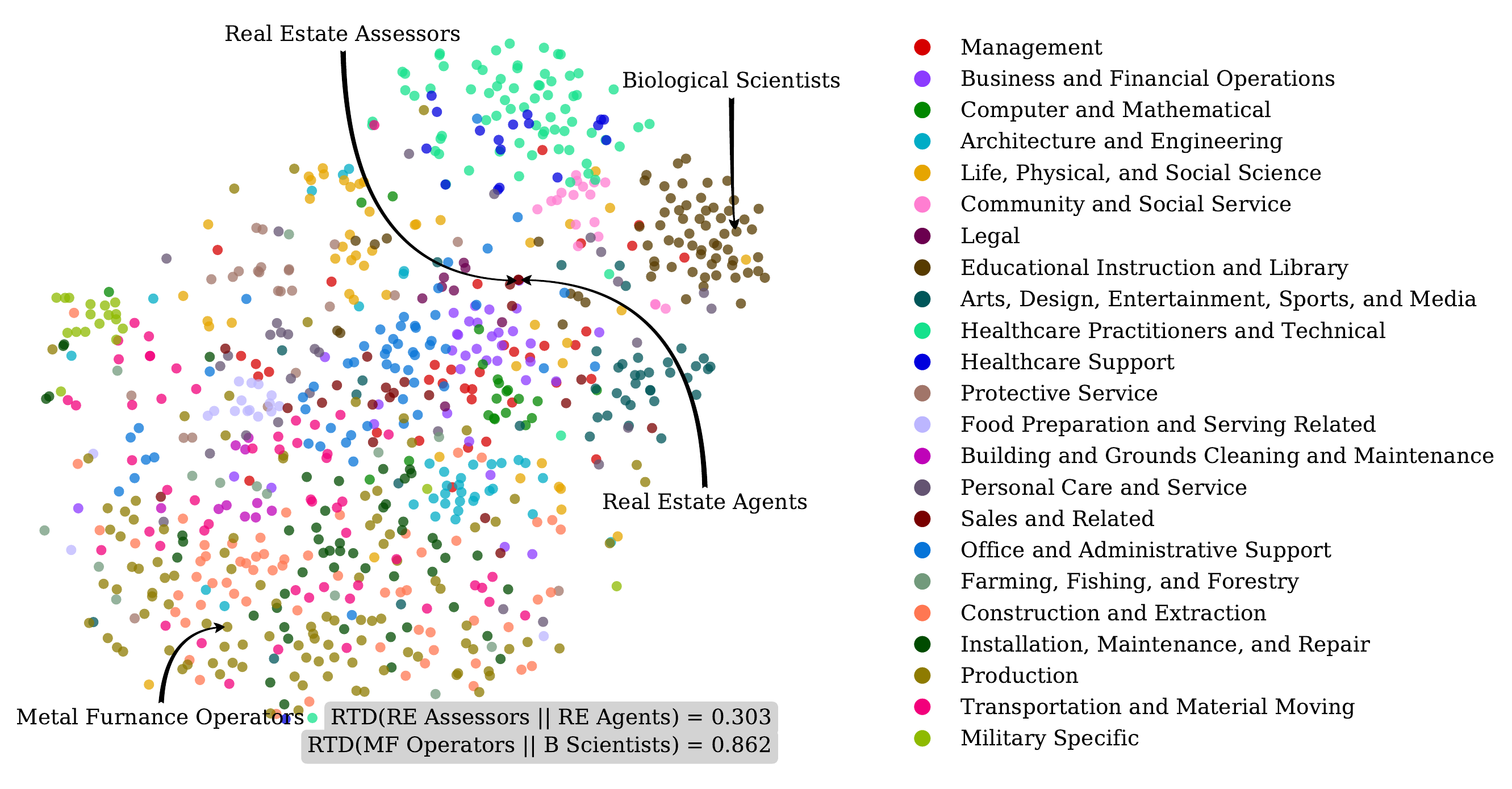}
    \caption{\textbf{Alternative measures of language divergence (i.e., RTD) also capture notable commonalities among detailed occupations.}
    Points are produced by tSNE embedding of language distribution rank-turbulence divergences.
    Each point corresponds to a detailed occupation and is colored by its corresponding major occupation.
    The location in the visualization is determined by an embedding created using the RTD divergence values as a pre-computed distance. 
    See Fig.~\ref{fig:visual_abstract}C for the JSD equivalent. 
   }
    \label{fig:tSNE_RTD_space}
\end{figure*}

In Fig.~\ref{fig:tSNE_RTD_space} to we present a tSNE embedding of occupations based on the pairwise RTD values for their language distributions (similar to the JSD version in Fig.~\ref{fig:visual_abstract}C).
Using the RTD measure, two of the most similar detailed occupations are \texttt{Appraisers and Assessors of Real Estate} and \texttt{Real Estate Sales Agents} (RTD = 0.303). 
On the other hand, two of the most different detailed occupations---as determined by RTD---are \texttt{Ophthalmologists, Except Pediatric} and \texttt{Cutting, Punching, and Press Machine Setters, Operators, and Tenders, Metal and Plastic} (RTD=0.862)

\clearpage

\subsection{Model feature descriptions}\label{sec:model_features}

\begin{table}[h!]
    \centering
    \begin{tabular}{|c|p{.6\linewidth}|}
    \hline
       \textbf{Feature name}  &  \textbf{Description}\\
       \hline
        JSD$_{\textnormal{gender}}$ & Jensen-Shannon divergence (JSD) values for language distributions of female and male works in detailed occupations.\\
        \hline
        JSD$_{\textnormal{major}}$ & JSD values for language distributions of detailed occupations and the language distribution of major occupation (with the distribution of the given detailed occupation removed from the major distribution). \\
        \hline
        w2v~bias &  Word-Embedding Association Test (WEAT) effect sizes. \\
        \hline
        State quotient & State employment quotients for 50 states and 6 territories---defined as the percentage of a state's work force employed in a detailed occupation divided by the national percentage of workers in that detailed occupation. \\
        \hline
        Female employment & Gender balance of female and male workers in the detailed occupation (Female~employment~=~$N_f / N_m$).
        \\
        \hline
        Major SOC FE & Fixed effects for SOC major occupation categories.
        \\
        \hline
        Year FE & Year fixed effects.
        \\
        \hline
    \end{tabular}
    \caption{Feature names and descriptions for the full model.}
    \label{tab:feature_names}
\end{table}

\subsection{Word-Embedding Association Test}
\label{sec:weat}
We use the methodology from Caliskan \textit{et al.} (2017) \cite{caliskan2017semantics} to measure the association of 1-grams from resumes with the concept of gender.
At its core, the Word-Embedding Association Test (WEAT) relies on cosine similarity scores between attribute words and target words.
Examples of attribute categories from Caliskan \textit{et al.} include pleasant vs. unpleasant, temporary vs. permanent, and male vs. female.
Examples of target categories include instruments vs. weapons, math vs. arts, and young vs. old people's names. 
In our case, the attribute words are male vs. female and the target words are the top 50 words contributing to JSD from the female and male language distributions (100 words total) for a detailed occupation.

The association is measured by comparing distributions for women and men with the gender attribute terms and looking at the standardized difference between the similarity scores. 
More specifically, the test statistic for each target word $w$ is defined by

\[
s(w, A, B) = \frac{\textnormal{mean}_{a\in A}
\textnormal{cos}(\overrightarrow{w}, \overrightarrow{a}) - 
\textnormal{mean}_{b\in B}
\textnormal{cos}(\overrightarrow{w}, \overrightarrow{b})}
{\textnormal{std\_dev}
_{x \in A\cup B } 
\textnormal{cos}
(\overrightarrow{w},\overrightarrow{x})
}
\]

The difference in $s(w, A, B)$ for all target words associated with the female and male distributions provides the final bias measure. 
This is defined by

\[ s(X,Y,A,B) = \sum_{x \in X} s(x, A, B) - \sum_{y \in y} s(y, A, B) \]

Where $X$ and $Y$ are our target terms from female and male resumes. $A$ and $B$ are attribute terms corresponding to female and male genders. 

We use the pretrained \texttt{word2vec-google-news-300} model from the original word2vec paper for all WEAT tests \cite{mikolov2013efficient}. 
While the model is older and far from state-of-the-art, the main point is to gain a measure of gender association from a word embedding space, a task that is well understood with an older model such as word2vec. 
Caliskan \textit{et al.} (2017) find similar results when using the GloVe \cite{pennington2014glove} word-embedding model.

\subsection{Skills embeddings}\label{sec:skills_embed_method}

We match text descriptions of positions from resumes with DWAs from O*NET using sentence transformers \cite{reimers-2019-sentence-bert}.
Sentence transformers are transformers-based language models (e.g., BERT) that are optimized for semantic similarity tasks.
Specifically we use the \texttt{all-mpnet-base-v2} model but will generically refer to it as SBERT.
SBERT allows us to establish a link between latent skills present in resumes and canonical skillsets from O*NET. 
In this we have SBERT embeddings of worker resumes, $\overrightarrow{r}$, and O*NET detailed workplace activities (DWAs), $\overrightarrow{a}$, for a given SOC.
More specifically, 
\[
    \overrightarrow{a_i} = \textnormal{mean}_{i\in soc}(\overrightarrow{\textnormal{activities}_i})
\]

To match resumes with SOCs using their associated DWAs, we take the maximum cosine similarity score between a resume embedding $\overrightarrow{r}$ and all pairwise comparisons with activities embeddings $\overrightarrow{a}$. 
Assigning resume $r_j$ to a detailed occupation with the following: 

$\argmax$

\[ SOC_j = \argmax_i (\textnormal{cos}( \overrightarrow{r_j}, \overrightarrow{a_i} )) \quad\quad, \forall i \]

The detailed occupation with a DWA cluster that is most similar to the resume embedding is assigned to the resume. 
We compare these results with the title-matched detailed occupations assigned by the procedure described in Sec.~\ref{subsec:jobtitleclass}.

\begin{figure*}[h!]
    \centering
   \includegraphics[width=\textwidth]{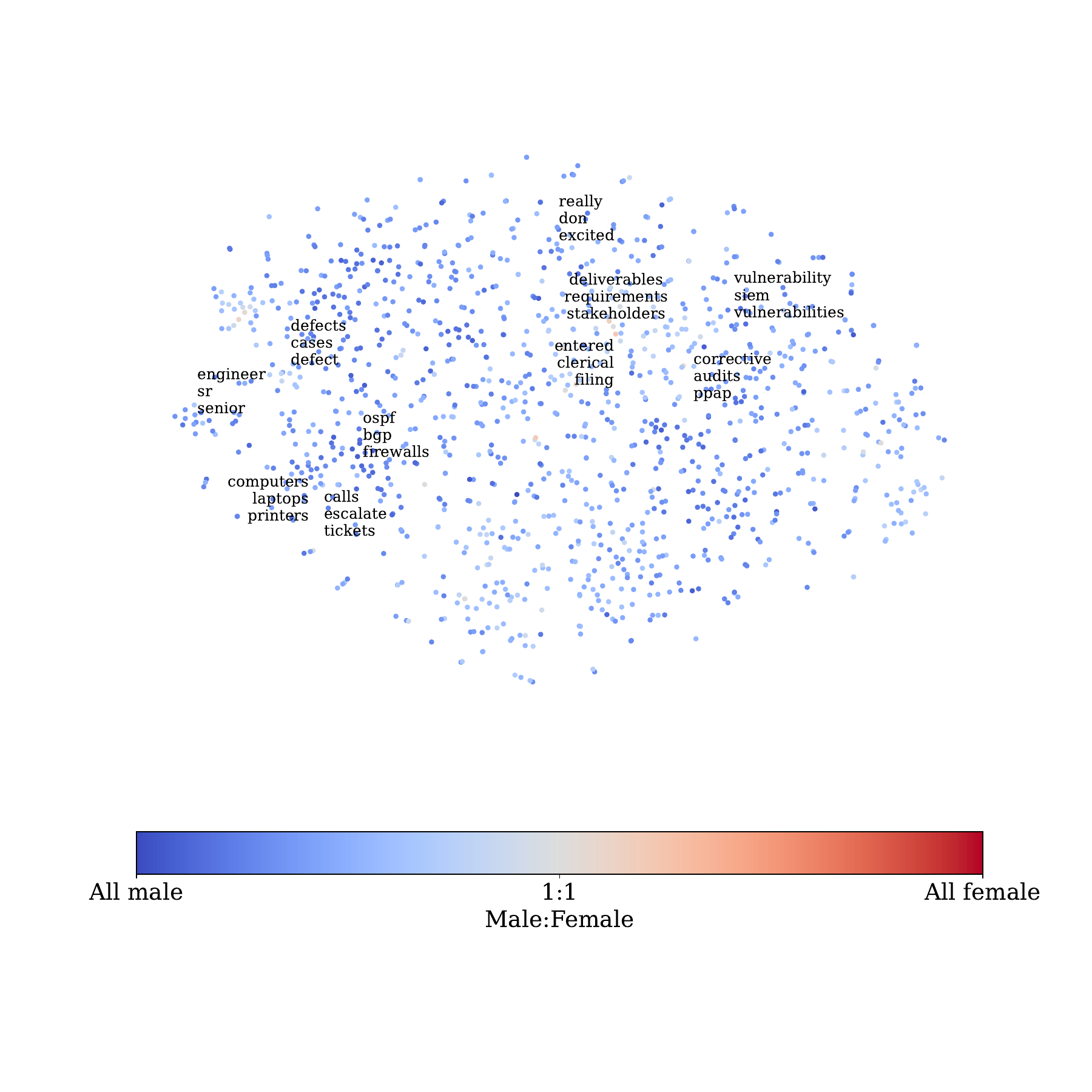}
    \caption{\textbf{tSNE embedding of top2vec topics for resumes in the Computer and Mathematical Occupations major SOC}.
    }
    \label{fig:tSNE_top2vec}
\end{figure*}

\begin{figure*}[h]
    \centering
   \includegraphics[width=\textwidth]{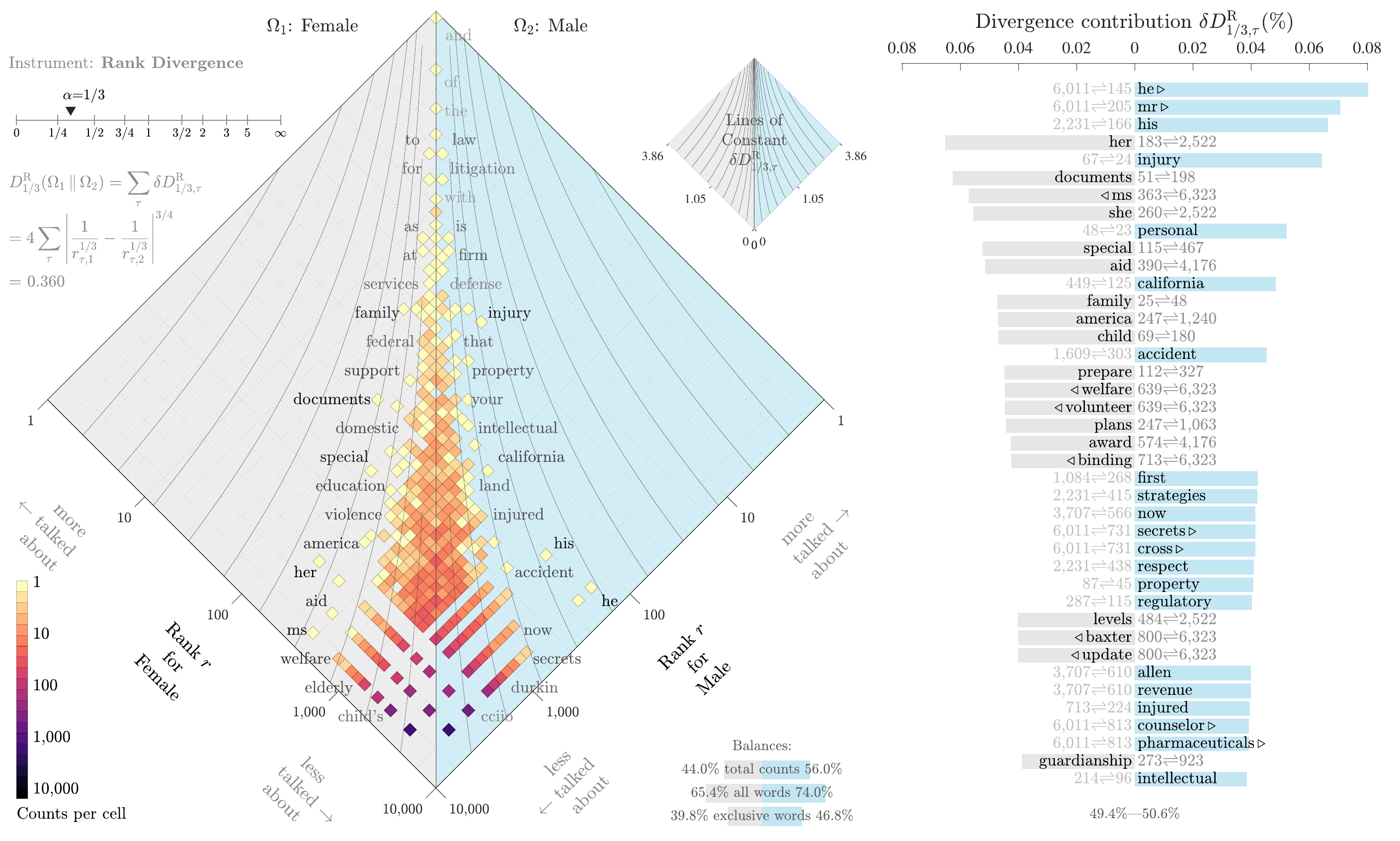}
    \caption{\textbf{Allotaxonograph \cite{dodds2020allotaxonometry} for female and male lawyers (detailed occupation).}
    The central diamond shaped plot shows a rank-rank histogram for 1-grams appearing in each gender's resume language distributions.
    The horizontal bar chart on the right shows the individual contribution of each 1-gram to the overall rank-turbulence divergence value ($D^\text{R}_{1/3}$). 
    The triangles to the right or left of $1$-grams indicate that the term is unique to that distribution. 
    The 3 bars under ``Balances'' represent the total volume of 1-gram occurring in each distribution, the percentage of all unique words we saw in each distribution, and the percentage of words that we saw in a distribution that were unique to that distribution.
    }
    \label{fig:allotaxonographsoftware}
\end{figure*}

\begin{figure*}[h!]
 \centering

\includegraphics[width=.8\columnwidth]{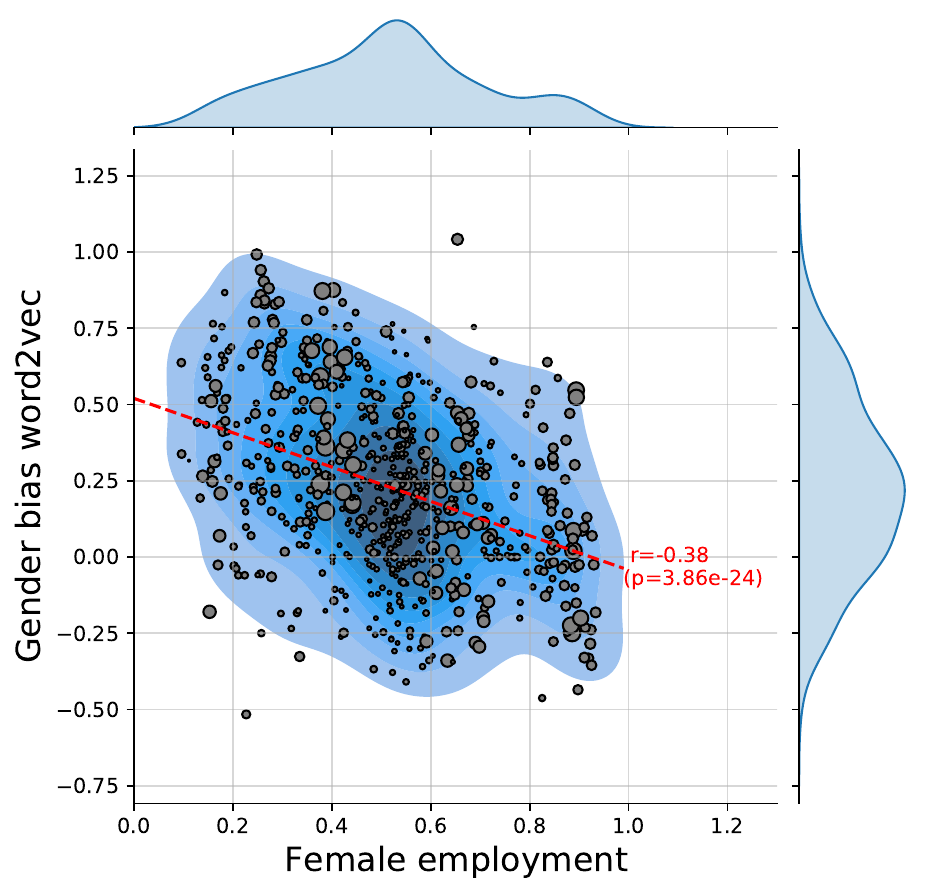}
    \caption{\textbf{Bias scores for word2vec-derived gender bias effect compared with proportion of workers who are female for detailed occupations.}}
    \label{fig:w2vempshare}

\end{figure*}

\begin{figure*}[h!]
    \centering
    \includegraphics[width=.8\columnwidth]{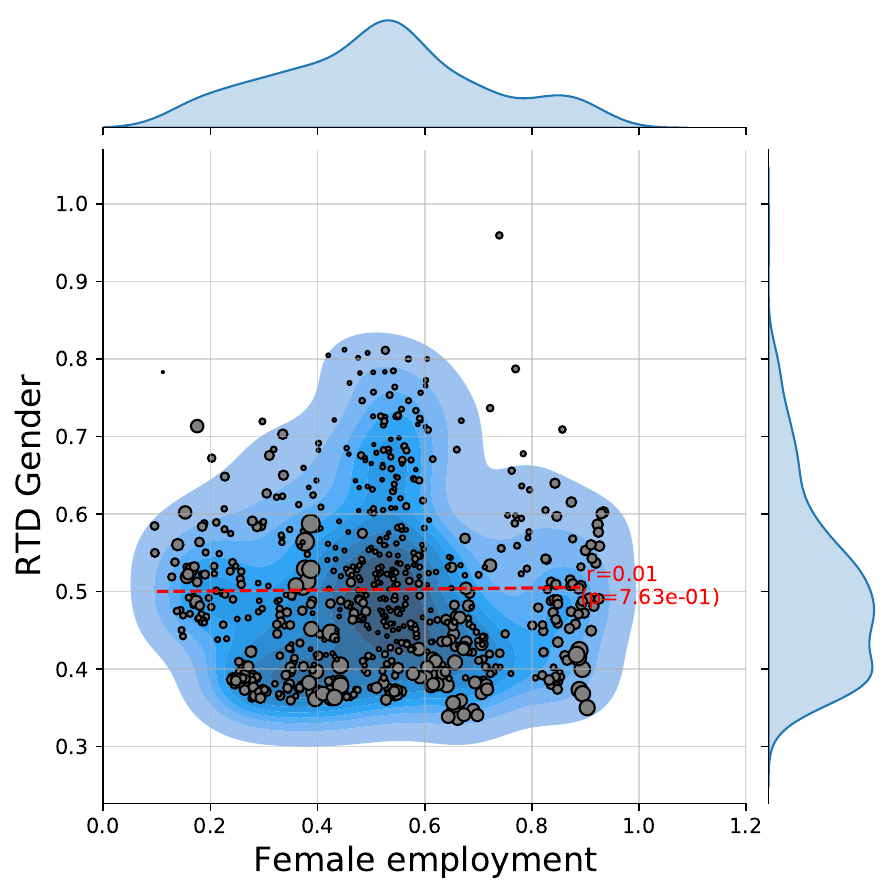}
    \caption{\textbf{Rank-turbulence divergence over female employment share}}
    \label{fig:rtdempshare}

\end{figure*}

\begin{figure*}[h!]
    \centering
    \includegraphics[width=.85\columnwidth]{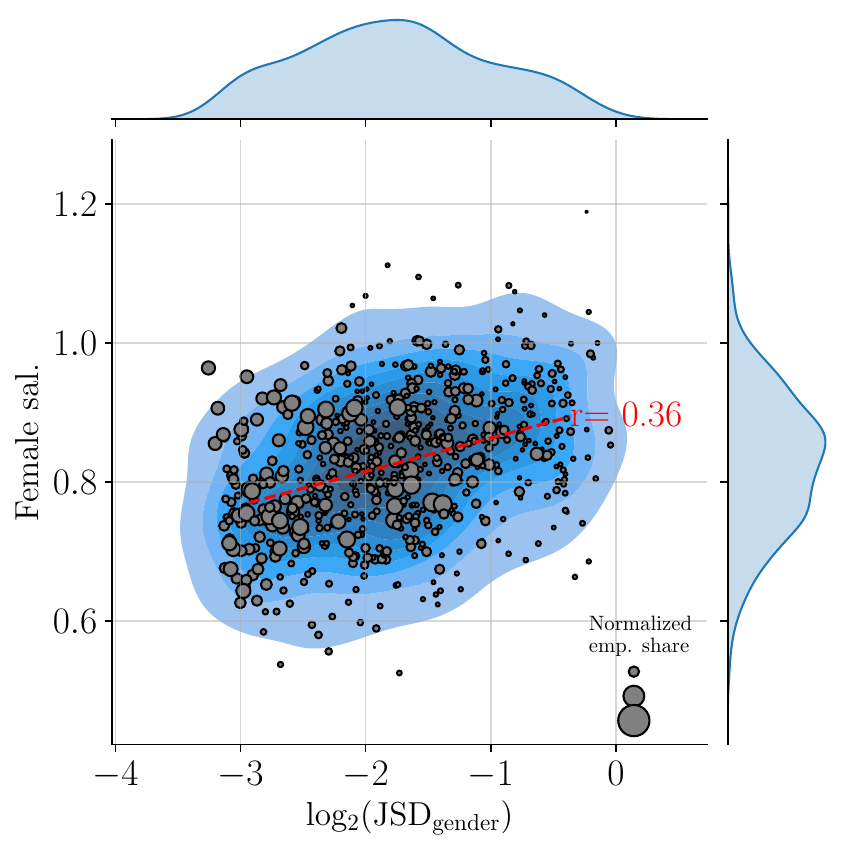}
    \caption{\textbf{Gender wage gap over log$_2$ transformed Jensen-Shannon divergence for female and male resume language distributions. }
    See \ref{fig:visual_abstract}B for linear JSD$_\textnormal{gender}$ version. }
    \label{fig:log2correlation_full}
\end{figure*}

\begin{figure*}
    \centering
    \includegraphics[width=.9\columnwidth]{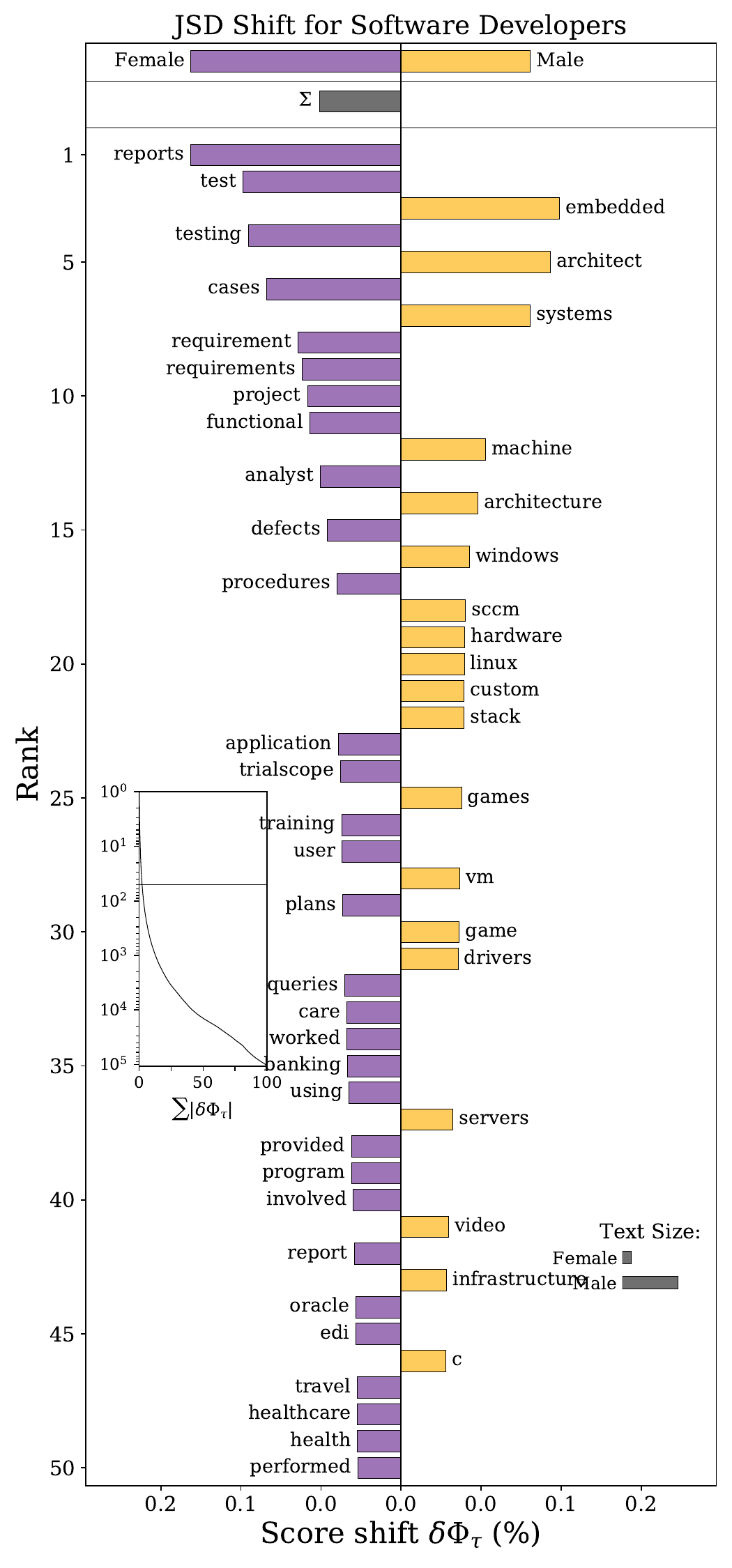}
    \caption{\textbf{Word shift \cite{gallagher2021generalized} for 1-gram contributions to JSD for female and male software developers. }
    }
    \label{fig:jsd_shift_sotware_full}
\end{figure*}

\begin{samepage}

\begin{table*}[h!]
    \centering
    \renewcommand{\arraystretch}{1.25}
    \tiny
    \begin{tabular}{M{2.5cm}|M{1.1cm}|M{1.1cm}|M{1.1cm}|M{1.1cm}|M{1.1cm}|M{1.1cm}|M{1.1cm}|M{1.1cm}|M{1.1cm}}
    \hline
        \multicolumn{10}{c}{\textbf{Dependent variable:}} \\
        \multicolumn{10}{c}{Female wage percentage} \\
        \hline
        \hline
Variable & Model 1 & Model 2 & Model 3 & Model 4 & Model 5 & Model 6 & Model 7 & Model 8 & Model 9 \\  \hline Female emp. & 0.081$^{***}$ (0.009) &  & 0.071$^{***}$ (0.009) & 0.078$^{***}$ (0.009) &  & 0.046$^{***}$ (0.011) & 0.059$^{***}$ (0.011) & 0.061$^{***}$ (0.011) & 0.061$^{***}$ (0.011) \\ log$_{2}$(JSD$_{\textnormal{gender}}$) &  & 0.048$^{***}$ (0.005) & 0.068$^{***}$ (0.008) &  &  &  & 0.051$^{***}$ (0.010) & 0.052$^{***}$ (0.011) &  \\ log$_{2}$(JSD$_{\textnormal{major}}$) &  &  & -0.039$^{***}$ (0.010) &  &  &  & -0.009$^{}$ (0.012) & -0.008$^{}$ (0.012) &  \\ log$_{2}$(RTD$_{\textnormal{gender}}$) &  &  &  & 0.105$^{***}$ (0.014) &  &  &  &  & 0.071$^{***}$ (0.020) \\ log$_{2}$(RTD$_{\textnormal{major}}$) &  &  &  & -0.027$^{}$ (0.021) &  &  &  &  & 0.044$^{}$ (0.035) \\ w2v bias gen. &  &  &  &  & -0.087$^{***}$ (0.013) &  &  & 0.006$^{}$ (0.014) &  \\ State quotient &  &  &  &  &  & Yes & Yes & Yes & Yes \\ Year FE &  &  &  &  &  & Yes & Yes & Yes & Yes \\ Major SOC FE &  &  &  &  &  & Yes & Yes & Yes & Yes \\ R$^{2}$ & 0.100 & 0.130 & 0.230 & 0.190 & 0.063 & 0.528 & 0.559 & 0.559 & 0.543 \\ Adj. R$^{2}$ & 0.099 & 0.129 & 0.226 & 0.186 & 0.061 & 0.460 & 0.493 & 0.492 & 0.475 \\
    \hline
    \hline
    \multicolumn{10}{c}{$p < 0.1^*,\ p < 0.01^{**},\ p < 0.001^{***}$} \\
    \hline
    \end{tabular}
\caption{\textbf{Ordinary least squares models predicting the wage percentage for women.}
Coefficient values are reported with standard errors in parentheses. 
State quotients, year fixed effects (FE) and Major SOC FE are reported as binary inclusion.
}
\label{tab:sal_share_model}
\end{table*}

\begin{table*}[h!]
    \centering
    \renewcommand{\arraystretch}{1.25}
        \tiny
    \begin{tabular}{M{2.5cm}|M{1.1cm}|M{1.1cm}|M{1.1cm}|M{1.1cm}|M{1.1cm}|M{1.1cm}|M{1.1cm}|M{1.1cm}|M{1.1cm}}
    \hline
        \multicolumn{10}{c}{\textbf{Dependent variable:}} \\
        \multicolumn{10}{c}{Female employment percentage} \\
        \hline
        \hline
Variable & Model 1 & Model 2 & Model 3 & Model 4 & Model 5 & Model 6 & Model 7 & Model 8 & Model 9 \\  \hline Female sal. & 1.237$^{***}$ (0.155) &  & 1.254$^{***}$ (0.160) & 1.298$^{***}$ (0.159) &  & 0.614$^{***}$ (0.156) & 0.792$^{***}$ (0.154) & 0.739$^{***}$ (0.145) & 0.750$^{***}$ (0.152) \\ log$_{2}$(JSD$_{\textnormal{gender}}$) &  & 0.043$^{*}$ (0.017) & 0.017$^{}$ (0.037) &  &  &  & -0.119$^{**}$ (0.044) & -0.120$^{**}$ (0.041) &  \\ log$_{2}$(JSD$_{\textnormal{major}}$) &  &  & -0.059$^{}$ (0.050) &  &  &  & -0.074$^{}$ (0.054) & -0.113$^{*}$ (0.051) &  \\ log$_{2}$(RTD$_{\textnormal{gender}}$) &  &  &  & -0.055$^{}$ (0.051) &  &  &  &  & -0.470$^{***}$ (0.062) \\ log$_{2}$(RTD$_{\textnormal{major}}$) &  &  &  & -0.232$^{**}$ (0.099) &  &  &  &  & -0.158$^{}$ (0.128) \\ w2v bias gen. &  &  &  &  & -0.589$^{***}$ (0.048) &  &  & -0.338$^{***}$ (0.047) &  \\ State quotient &  &  &  &  &  & Yes & Yes & Yes & Yes \\ Year FE &  &  &  &  &  & Yes & Yes & Yes & Yes \\ Major SOC FE &  &  &  &  &  & Yes & Yes & Yes & Yes \\ R$^{2}$ & 0.100 & 0.007 & 0.104 & 0.118 & 0.189 & 0.589 & 0.614 & 0.649 & 0.631 \\ Adj. R$^{2}$ & 0.099 & 0.005 & 0.100 & 0.114 & 0.188 & 0.529 & 0.557 & 0.596 & 0.576
\\
    \hline
    \hline
    \multicolumn{10}{c}{$p < 0.1^*,\ p < 0.01^{**},\ p < 0.001^{***}$} \\
    \hline
    \end{tabular}
\caption{\textbf{Ordinary least squares models predicting the employment share percentage for women.}
Coefficient values are reported with standard errors in parentheses. 
State quotients, year fixed effects (FE) and Major SOC FE are reported as binary inclusion.
}
\label{tab:emp_share_model}
\end{table*}
\end{samepage}

\clearpage
\vspace{0cm}

\begin{table}
    \centering
    \begin{tabular}{l|r}
    \hline 
       Variable  & VIF  \\
       \hline
       $\textnormal{JSD}_\textnormal{gender}$   & 3.48 \\
       $\textnormal{JSD}_\textnormal{major} $   & 3.48 \\
       Female emp.                              & 1.01 \\
       \hline 
    \end{tabular}
    \caption{\textbf{Variance inflation factor for gender language distribution divergence, detailed-major occupation language distribution divergence, and female employment share.}}
    \label{tab:VIF}
\end{table}

\begin{samepage}
\subsection{Data sources}
\begin{itemize}
    \item US Bureau of Labor Statistics (BLS) standard occupation classification (SOC) \cite{blsStandardOccupational}
    \item FutureFit AI
    \item BLS weekly median earnings by gender \cite{blsTables}
    \item US Census MSA data
    \item OES national occupation employment by state \cite{bls2021National}
\end{itemize}
\end{samepage}

\end{document}